\documentclass[journal,twoside,web]{ieeecolor}

\usepackage{tuffc}
\usepackage{cite}
\usepackage{amsmath,amssymb,amsfonts}
\usepackage{graphicx}
\usepackage{textcomp}
\usepackage{import}
\usepackage{wrapfig,colortbl}
\definecolor{abstractbg}{rgb}{1,0.969,0.914}
\graphicspath{{figures/}}
\DeclareGraphicsExtensions{.pdf,.jpeg,.png}
\usepackage{hyperref}
\usepackage{url}
\usepackage{color}
\usepackage{pifont}
\usepackage{booktabs}
\usepackage{multirow}
\usepackage[utf8]{inputenc}
\usepackage[T1]{fontenc}
\usepackage{authblk}
\usepackage{stfloats}
\usepackage{xspace}

\usepackage{caption}
\usepackage{algorithm}
\usepackage{algpseudocode}

\newcommand{\nnunet}{\mbox{nnU-Net}\xspace}
\newcommand{\pdnet}{\mbox{PDNet}\xspace}
\newcommand{\oripdnet}{\mbox{OriPDNet}\xspace}
\newcommand{\trans}{\mbox{BATFormer}\xspace}
\newcommand{\DeAN}{\mbox{DeAN}\xspace}

\newcommand{\ie}{\mbox{\textit{i.e.}}\xspace}

\newcommand{\etal}{\mbox{\textit{{et al.}}}\xspace}

\newcommand{\blue}[1]{\textcolor{blue}{#1}}
\definecolor{pink}{rgb}{0.858, 0.188, 0.478}
\definecolor{dark_yellow}{rgb}{1, 0.76, 0}
\definecolor{orange}{rgb}{1.0, 0.5, 0.0}

\definecolor{cyan}{rgb}{0.113, 0.776, 0.686}
\definecolor{brown}{rgb}{0.741, 0.552, 0.180}
\definecolor{mygreen}{rgb}{0.0, 0.5, 0.0}
\definecolor{eucalyptus}{rgb}{0.3125, 0.7891, 0.6758}

\newcommand{\sota}{\mbox{state-of-the-art}\xspace}

\def\BibTeX{{\rm B\kern-.05em{\sc i\kern-.025em b}\kern-.08em
    T\kern-.1667em\lower.7ex\hbox{E}\kern-.125emX}}
\usepackage{xcolor}
\usepackage{soul}

\begin{document}
\bstctlcite{IEEEexample:BSTcontrol}

\title{Phase Unwrapping of Color Doppler Echocardiography using Deep Learning}

\author{
Hang Jung~Ling, 
Olivier~Bernard,
Nicolas~Ducros, and
Damien~Garcia
\thanks{Manuscript received 14 April 2023; revised 5 June 2023; accepted 20 June 2023.This work was supported in part by the MEGA Doctoral School (ED 162); in part by the French National Research Agency (ANR) through the “4D-iVFM” Project under Grant ANR-21-CE19-0034-01; in part by LABEX PRIMES, Université de Lyon, under Grant ANR-11-LABX-0063; in part by LABEX CELYA within the program “Investissements d’Avenir” operated by the French ANR under Grant ANR-10-LABX-0060; and in part by GENCI-IDRIS (HPC resources) under Grant 2022- [AD010313603]. (Corresponding author: Hang Jung Ling.)}
\thanks{Hang Jung Ling, Olivier Bernard, Nicolas Ducros and Damien Garcia are with CREATIS, CNRS UMR 5220, INSERM U1294, INSA Lyon, University of Lyon 1, 69100 Villeurbanne, France (e-mail: hang-jung.ling@creatis.insa-lyon.fr; damien.garcia@creatis.insa-lyon.fr).}
\thanks{Nicolas Ducros is also with Institut Universitaire de France (IUF), France.}}

\IEEEtitleabstractindextext{%
\fcolorbox{abstractbg}{abstractbg}{%
\begin{minipage}{\textwidth}\rightskip2em\leftskip\rightskip\bigskip
\begin{wrapfigure}[20]{r}{3in}%
\vspace{-1.1pc}\hspace{-3pc}\includegraphics[width=2.9in]{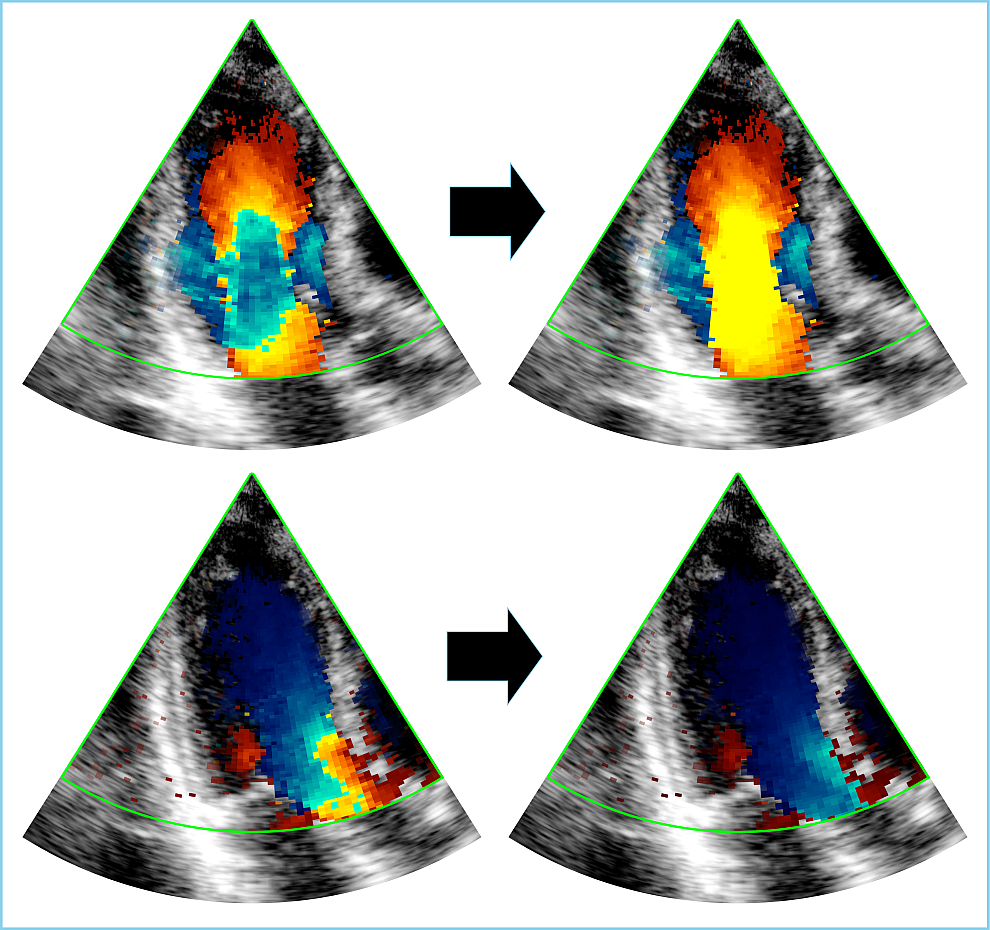}
\end{wrapfigure}%
\begin{abstract}
Color Doppler echocardiography is a widely used non-invasive imaging modality that provides real-time information about the intracardiac blood flow. In an apical long-axis view of the left ventricle, color Doppler is subject to phase wrapping, or aliasing, especially during cardiac filling and ejection. When setting up quantitative methods based on color Doppler, it is necessary to correct this wrapping artifact. We developed an unfolded primal-dual network to unwrap (dealias) color Doppler echocardiographic images and compared its effectiveness against two state-of-the-art segmentation approaches based on \nnunet and transformer models. We trained and evaluated the performance of each method on an in-house dataset and found that the \nnunet-based method provided the best dealiased results, followed by the primal-dual approach and the transformer-based technique. Noteworthy, the primal-dual network, which had significantly fewer trainable parameters, performed competitively with respect to the other two methods, demonstrating the high potential of deep unfolding methods. Our results suggest that deep learning-based methods can effectively remove aliasing artifacts in color Doppler echocardiographic images, outperforming \DeAN, a state-of-the-art semi-automatic technique. Overall, our results show that deep learning-based methods have the potential to effectively preprocess color Doppler images for downstream quantitative analysis.
\end{abstract}

\bigskip
\end{minipage}}}

\maketitle
\IEEEpeerreviewmaketitle

\section{Introduction}
\label{sec:introduction}

\IEEEPARstart{C}{olor} Doppler ultrasound is a widely accepted clinical imaging modality for non-invasive, real-time analysis of cardiovascular blood flow. While two-dimensional color Doppler is commonly used for qualitative mapping of flow characteristics, its applications for quantitative analysis are limited. Common cardiovascular applications of color Doppler include detection of valvular diseases \cite{zoghbi2017} and septal defects \cite{hanumansetty2022}, or guiding the positioning of the pulsed-wave sample volume for spectral Doppler \cite{revzin2019}. Among methods to make color Doppler quantitative, Vector Flow Mapping has been introduced for intracardiac flow dynamics. This method allows the computation of 2D or 3D intraventricular velocity vector maps from color Doppler fields, using a physically constrained optimization approach \cite{phymedbio_vixege_2021,phymedbio_vixege_2022}. Intracardiac vector flow mapping from color Doppler requires two prerequisite steps: 1) delineation of the endocardial inner wall, and 2) correction of wrapped (aliased) Doppler regions.

With respect to the first step, Painchaud \etal \cite{painchaud2022} recently introduced a 2D+time deep learning architecture to enforce temporal consistency and smoothness from one frame to the next. The second step of correction is necessary due to the occurrence of aliasing, which is an artifact resulting from insufficient slow-time sampling. This issue arises when the pulse repetition frequency (PRF) is unable to capture high axial velocities effectively. This causes the Doppler velocity to be wrapped to the opposite side of the Doppler spectrum when its absolute value exceeds the Nyquist velocity. Experienced clinicians can easily identify zones of aliasing in most color Doppler images, where the color-coded velocities shift from red to blue or vice versa. Aliasing can be removed in Doppler echocardiography by designing multi-PRF sequences, as described by Posada \etal \cite{tmi_posada_2016}. However, this approach requires control of the ultrasound machine and is primarily suitable for high-frame-rate echocardiography. When clinical scanners are used, aliasing must be corrected by post-processing the color Doppler fields. While a number of unwrapping algorithms have been proposed for dealiasing data maps in atmospheric science,  geodesy, and optical interferometry \cite{josaa_ghiglia_1994,josaa_shanker_2010,martinez2017}, this problem has received less attention in color Doppler imaging. 

Inspired by traditional radar approaches, Muth \etal \cite{media_muth_2011} developed a segmentation-based method for color Doppler dealiasing using statistical region merging, called \DeAN. This unsupervised method uses a scalar hyperparameter to control the segmentation process. An optimal parameter was determined from a supervised analysis of 50 color Doppler data. However, it turns out that the \DeAN method fails in difficult cases as shown in Fig. 11 in \cite{media_muth_2011} and that supervised corrections are still necessary in some situations. With the goal of developing imaging tools that quantify blood flow from color Doppler, we propose a deep learning (DL) approach to correct the aliased areas of echocardiographic color Doppler maps. DL has been proposed for color Doppler dealiasing in vascular flow imaging by Nahas \etal \cite{uffc_nahas_2020}. Their approach aimed to solve the double aliasing problem using two U-Nets. The first U-Net detected the presence of single aliasing while the second U-Net was trained to identify and segment double-aliased pixels. They evaluated the performance of their model by training it with different types of ultrasound information. They found that the model trained with a combination of Doppler frequency, power, and bandwidth performed the best for dealiasing in the femoral bifurcation.

In our work, we focused on Doppler echocardiography. In contrast to vascular flow imaging, cardiac color flow imaging can be subject to substantial clutter signals originating from the myocardium and tending to spread the aliased patterns. With the goal of proposing a robust DL method that correctly handles aliasing in most situations, we developed and compared several architectures. Our main contributions are:

\begin{enumerate}
    \item  We designed a primal-dual network based on the idea of deep unfolding, and compared it with \sota DL segmentation methods and \DeAN.
    \item We used a private color Doppler echocardiographic dataset acquired in apical three-chamber view (45 patients, 1,338 aliased and 2,379 non-aliased frames) to train the neural networks and analyze their performance.
    \item We investigated the value of adding Doppler power as input information to improve dealiasing.
    \item We introduced a data augmentation strategy that generates synthetic aliasing, which solved the class imbalance problem and improved dealiasing performance on difficult color Doppler images.
\end{enumerate}

\begin{table*}[!t]
\arrayrulecolor{subsectioncolor}
\setlength{\arrayrulewidth}{1pt}
{\sffamily\bfseries\begin{tabular}{lp{6.75in}}\hline
\rowcolor{abstractbg}\multicolumn{2}{l}{\color{subsectioncolor}{\itshape
Highlights}{\Huge\strut}}\\
\rowcolor{abstractbg}$\bullet$ & Our deep-unfolding-based primal-dual network (PDNet) incorporated the forward operator as prior information and had only 0.03M parameters.\\

\rowcolor{abstractbg}$\bullet${\large\strut} & Our deep learning (DL) models outperformed a state-of-the-art non-DL approach in phase unwrapping of color Doppler echocardiography, with \nnunet being the best candidate, followed by \pdnet with 233 times fewer parameters.\\

\rowcolor{abstractbg}$\bullet${\large\strut} & Automatic and accurate color Doppler echocardiographic phase unwrapping ensures correct visualization and enables quantification of intracardiac blood flow.\\
[2em]\hline
\end{tabular}}
\setlength{\arrayrulewidth}{0.4pt}
\arrayrulecolor{black}
\end{table*}

\section{Methodology} \label{sec:methodology}
Aliasing artifacts occur when axial blood speeds (velocity magnitudes) exceed the Nyquist velocity $V_N$. The acquired Doppler velocity $V_D$ can be written as a function of the unwrapped or alias-free Doppler velocity $V_u$ as follows:
\begin{equation} \label{eq:unwrapping}
    V_D = V_u - 2 \times n_N\,V_N,
\end{equation}
where $n_N$ is an integer called the Nyquist number, which represents the number of times the signal wraps around the Nyquist limit. 
The Nyquist number reads (see \cite{tmi_posada_2016} for the demonstration)
\begin{equation} \label{eq:Nyquist_number}
    n_N = \text{floor}\left(\frac{V_u+V_N}{2V_N}\right).
\end{equation}

Except for highly turbulent flows that may occur in transvalvular or transseptal jets, there is no multiple aliasing in the adult left ventricle scanned in the apical long-axis view, \ie, the integer $n_N$ belongs to $\{-1, 0, 1\}$.
Indeed, in adult echocardiography with a 3 MHz phased array, Nyquist velocities typically range from 0.55 to 0.7 m/s. Thus, single (\ie, $n_N = -1\,\text{or}\,1$) or no (\ie, $n_N = 0$) wrapping occurs as long as the actual blood speed is less than 1.65-2.1 m/s (see \eqref{eq:forward_velocity}). It follows that double aliasing does not occur in the left ventricle in most situations without valvular disease or cardiac shunt.
Equations \eqref{eq:unwrapping} and \eqref{eq:Nyquist_number} can be rewritten to express $V_D$ as a wrapped version of $V_u$:
\begin{equation}\label{eq:wrapping}
    V_D = K(V_u) = (V_u+V_N)\,\text{mod}\,(2V_N) - V_N,
\end{equation}
where \text{mod} is the modulo operation. In particular, for $n_N\in\{-1,0,1\}$, the wrapping function $K$ becomes
\begin{equation}
    V_D = K(V_u) =
    \small{
    \begin{cases}
    V_u - 2 V_N & \text{if } V_N < V_u < 3\,V_N\\
    V_u & \text{if } - V_N\le V_u \le V_N\\
    V_u + 2 V_N  & \text{if } -3\,V_N < V_u < -V_N
    \label{eq:forward_velocity}
    \end{cases}.
    }
\end{equation}

\begin{figure*}[tp]
    \centering
    \includegraphics[width=0.78\textwidth]{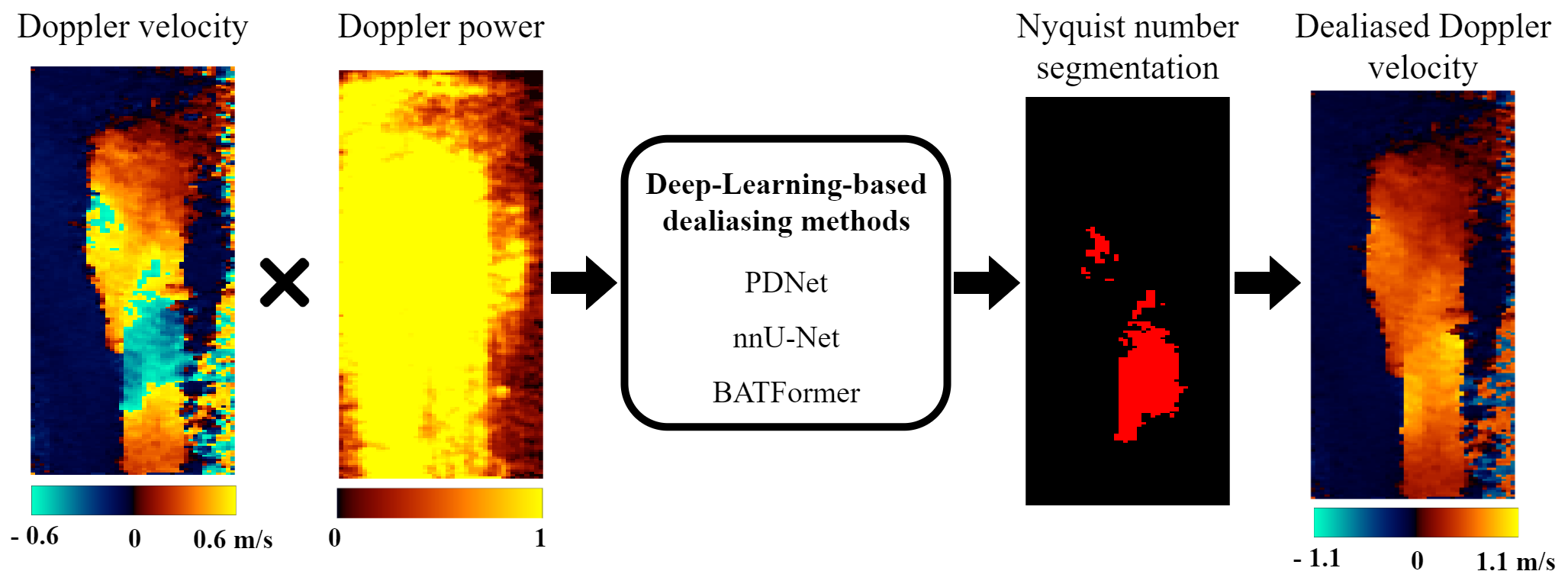}
    \caption{Pipeline of the deep learning-based methods for color Doppler dealiasing.}
    \label{fig:main_illustration}
\end{figure*}

This representation implies that the dealiasing problem can be approached in two different ways: \emph{i)} a problem that inverts the wrapping function \eqref{eq:wrapping} and recovers $V_u$ from the Doppler velocities $V_D$ by changing absolute jumps greater than $V_N$ to their $2{\times}V_N$ complement; \emph{ii)} a multi-class segmentation approach that assigns a Nyquist number $n_N$ \eqref{eq:Nyquist_number} to each pixel of the input image then computes the actual unwrapped velocities using (\ref{eq:unwrapping}). We investigated three deep learning (DL) models for dealiasing color Doppler. The first method was derived from the deep unfolding/unrolling framework and solved the inverse problem defined in \eqref{eq:wrapping} to estimate the actual velocities.
We faced a nonlinear inverse problem on non-trivial data, whose solution may contain phase jumps at the blood/myocardium interfaces. Unrolled methods are well suited for solving inverse problems. Primal-dual optimization, on the other hand, is useful for nonlinear problems. For these reasons, we tested a learned primal-dual algorithm inspired by Adler \etal \cite{tmi_adler_2018}, as described in the following subsection. The other two methods were \sota networks that we have adapted for determining Nyquist numbers in color Doppler images. Fig. \ref{fig:main_illustration} illustrates the pipeline used for all three methods, whose input data were the Doppler velocity multiplied by the Doppler power before scan-conversion.

\subsection{\oripdnet: A Primal-Dual-based Deep Unfolding Network to Solve Inverse Problems
\label{sec:oripdnet}}

To solve our nonlinear inverse problem \eqref{eq:wrapping}, we used \oripdnet (refer to Fig. 2 in \cite{tmi_adler_2018} for the network architecture), a deep unfolding network based on a primal-dual optimization scheme \cite{tmi_adler_2018}. Given a general inverse problem aiming to obtain the solution $f$ from the measurement $g$ with the forward operator $K$:
\begin{equation}
    g = K(f),
    \label{eq:general_inverse_problem}
\end{equation}
the outline of \oripdnet to solve this problem is presented in Algorithm \ref{alg:original_PDNet}.

\begin{algorithm}[h!]
\caption{\oripdnet: Original primal-dual network}
\label{alg:original_PDNet}
\begin{algorithmic}
\State Initialize $\boldsymbol{f_0}$, $\boldsymbol{h_0}=[\textbf{0},\textbf{0},\textbf{0},\textbf{0},\textbf{0}] \in \mathbb{R}^{M \times N \times 5}$
\For {$i=1,\cdots,I$}
\State $\boldsymbol{h_i} \leftarrow \Gamma_{\theta^d_i}(\boldsymbol{h_{i-1}},K(\boldsymbol{f^{(2)}_{i-1}}), \boldsymbol{g})$

\State $\boldsymbol{f_i} \leftarrow \Lambda_{\theta^p_i}(\boldsymbol{f_{i-1}},[\partial K(\boldsymbol{f^{(1)}_{i-1})}]^*\, (\boldsymbol{h^{(1)}_i}))$

\EndFor
\State return $\boldsymbol{f^{(1)}_{I}}$ 
\end{algorithmic}
\end{algorithm}

\oripdnet involves several variables and operators, including the forward operator $K$, the adjoint of its Fr\'echet derivative $[\partial K]^*$, the input measured data $g$, the primal and dual variables $f_i$ and $h_i$, and the learned primal and dual proximal operators $\Lambda_{\theta^p_i}$ and $\Gamma_{\theta^d_i}$. Convolutional layers are used to learn these proximal operators. The hyperparameter $I$, which determines the number of iterations, requires careful tuning for each specific problem.
The primal and dual variables, $f_i$ and $h_i$, are initialized then iteratively updated using the learned primal and dual proximal operators, $\Lambda_{\theta^p_i}$ and $\Gamma_{\theta^d_i}$. The solution to the inverse problem \eqref{eq:general_inverse_problem} is obtained by extracting the first element of the primal variables, $f_I^{(1)}$. 
In \cite{tmi_adler_2018}, the authors recommended setting the dimension of the primal and dual spaces to five as the best compromise between memory usage and reconstruction quality, \ie, $f_i=[f_i^{(1)}, f_i^{(2)}, f_i^{(3)}, f_i^{(4)}, f_i^{(5)}] \in \mathbb{R}^{M \times N \times 5}$ and $h_i=[h_i^{(1)}, h_i^{(2)}, h_i^{(3)}, h_i^{(4)}, h_i^{(5)}] \in \mathbb{R}^{M \times N \times 5}$, where $(M \times N)$ is the size of the input data. We conducted preliminary testing and validated the use of five spaces, in accordance with their suggestion.

\subsection{\pdnet: A Deep Unfolding Network for Color Doppler Dealiasing}
\label{sec:deepunfolding}
To deal with our specific inverse problem \eqref{eq:wrapping} for color Doppler dealiasing, we adapted \oripdnet. The modified version was named as \pdnet and summarized in Algorithm \ref{alg:proposed_PDNet}, with the main changes highlighted in blue.

\begin{algorithm}[h!]
\caption{\pdnet: Proposed primal-dual network}
\label{alg:proposed_PDNet}
\begin{algorithmic}
\State Initialize $\boldsymbol{\blue{V_0}}$, $\boldsymbol{h_0=[0,0,0,0,0]} \in \mathbb{R}^{M \times N \times 5}$
\For {$i=1,\cdots,I$}
\State $\boldsymbol{h_i} \leftarrow \blue{\Gamma_{\theta^d}}(\boldsymbol{h_{i-1}},\blue{K}(\boldsymbol{\blue{V^{(2)}_{i-1}}}),\boldsymbol{\blue{V_D}})$
\State $\boldsymbol{\blue{V_i}} \leftarrow \blue{\Lambda_{\theta^p}}(\boldsymbol{\blue{V_{i-1}}}, \boldsymbol{h^{(1)}_i})$
\EndFor
\State return $\blue{C_{\theta}}(\boldsymbol{\blue{V^{(1)}_{I}}})$ \\ \\

with $\boldsymbol{V_u} = \boldsymbol{\blue{V^{(1)}_{I}}}$ and $\boldsymbol{n_N} = \blue{C_{\theta}}(\boldsymbol{\blue{V^{(1)}_{I}}})$.
\end{algorithmic}
\end{algorithm}

Specifically, we defined the forward operator $K$ as a wrapping function given by (\ref{eq:wrapping}). Despite the discontinuity of this function at each $V = V_N \pm 2k V_N$ (with $k\in\mathbb{Z}^*$), its derivative was an identity function, i.e., $\partial K(V)=\text{id}$. Thus, its adjoint  $[\partial K(V)]^*$ was also an identity function. Unlike the original approach (see Algorithm \ref{alg:original_PDNet}), we used the same feature maps for each iteration of the main loop, which significantly reduced the number of parameters to learn (30,000 instead of 30,000$\times I$, with $I$ being the number of iterations) while maintaining the same accuracy. We made this change to avoid training instabilities that we observed while experimenting with \oripdnet. We also added a convolutional layer $C_{\theta}$ at the end of the network to output the Nyquist number from the estimated velocities $V_I^{(1)}$. The main reason for this was to avoid non-integer Nyquist numbers due to the regressed velocities. For a fair comparison between \pdnet and \oripdnet, the same convolution layer $C_{\theta}$ was also applied to the output of \oripdnet.

\subsection{Segmentation Networks for Color Doppler Dealiasing}
\label{sec:segmentation_dealiasing}

\begin{table*}[tp]
	\centering
	\caption{Main configurations of the three methods evaluated in this study. \textit{Lowest resolution}: Size of the lowest resolution of feature maps in pixels. \textit{Down. scheme}: Downsampling scheme. \textit{Up. scheme}: Upsampling scheme. \textit{Optimization scheme}: Optimizer + initial learning rate (+ learning rate scheduler used). \textit{\# param.}: Number of trainable parameters.}
	\smallskip
	\begin{tabular*}{0.95\textwidth}
		{c c c c c c c c c c}        
		\toprule
		\multirow{2}*{Methods} & \multicolumn{1}{c}{Number of} & \multicolumn{1}{c}{Lowest} & \multicolumn{1}{c}{Down.} & \multicolumn{1}{c}{Up.} & \multicolumn{1}{c}{Normalization} & \multicolumn{1}{c}{Batch} & \multicolumn{1}{c}{Optimization} & \multicolumn{1}{c}{Loss} & \multirow{2}*{\# param.} \\
		& \multicolumn{1}{c}{feature maps} & \multicolumn{1}{c}{resolution} & \multicolumn{1}{c}{scheme} & \multicolumn{1}{c}{scheme} & \multicolumn{1}{c}{scheme} & \multicolumn{1}{c}{size} & \multicolumn{1}{c}{scheme} & \multicolumn{1}{c}{function} &  \\
		\midrule
		PDNet & $32\rightarrow 32 \rightarrow 5$ & 192$\times$40 & - & - & - & 4 &
		\begin{tabular}{@{}c@{}} ADAM\cite{iclr_adam_2015} + \\ 0.001 + cosine \\ annealing\end{tabular}
		& 
		\begin{tabular}{@{}c@{}} Cross entropy \\ + Dice \end{tabular}
		& 0.03M \\
		
		nnU-Net & $32\downarrow480\uparrow32$ & 12$\times$5 & 
		\begin{tabular}{@{}c@{}} Stride \\ pooling\end{tabular}
		&
		\begin{tabular}{@{}c@{}} Transposed \\ conv.\end{tabular}
		&
		\begin{tabular}{@{}c@{}} Instance \\ norm.\end{tabular}
		& 4 &
		\begin{tabular}{@{}c@{}} SGD + 0.01 + \\ polynomial \\ decay\end{tabular}
		& 
		\begin{tabular}{@{}c@{}} Cross entropy \\ + Dice \end{tabular}
		& 7M \\
		
		\trans & $16\downarrow 256 \uparrow 16$ & 16$\times$16 & 
		\begin{tabular}{@{}c@{}} Max \\ pooling\end{tabular}
		&
		\begin{tabular}{@{}c@{}} 2 $\times$ 2 \\ repeats\end{tabular}
		&
		\begin{tabular}{@{}c@{}} Batch \\ norm.\end{tabular}
		& 4 &
		\begin{tabular}{@{}c@{}} ADAM + \\ 0.001\end{tabular}
		& 
		\begin{tabular}{@{}c@{}} Cross entropy \\ + Dice + \\ smooth L1\end{tabular}
		& 1.2M \\ 
		\bottomrule
	\end{tabular*}
	\label{tab:network_architectures}
\end{table*}

\nnunet is currently one of the best performing approaches for medical image segmentation \cite{isensee_nnu-net_2021}. This model is based on the \mbox{U-Net} architecture and implements several successful DL tricks, such as automatic hyperparameter search of the \mbox{U-Net} architecture to increase accuracy, a patch-wise approach to preserve image resolution, a deep supervision strategy to maintain accuracy at all scales, and data augmentation during both training and inference to enforce generalization. In this study, we addressed the dealiasing of color Doppler as a 3-class segmentation problem with \nnunet, where each class corresponded to a Nyquist number $n_N\in\{-1,0,1\}$. Our network included four stages in the encoder/decoder parts and had an input size of 192 $\times$ 40 pixels, which was the median image size of our dataset. Table \ref{tab:network_architectures} provides more details about the architecture and the training scheme (see Fig. C.1 in \cite{isensee_nnu-net_2021} for an illustration of the \nnunet architecture).

Recently, transformer-based approaches have been shown to outperform the \nnunet model in some medical challenges \cite{Antonelli2022}. These models attempt to solve the segmentation problem in a different way, using attention mechanisms with receptive fields that cover the entire image. Among the best-performing models, we chose to train the \trans architecture \cite{arxiv_lin_2022} for the color Doppler dealiasing task, using a segmentation-based technique. This model employs a multiscale approach based on a \mbox{U-Net} architecture, with transformer blocks added to the decoder. This strategy results in an efficient and lightweight architecture (1.2M parameters) that is suitable for learning from small to medium-sized datasets. The main configurations of \trans are listed in Table \ref{tab:network_architectures}, and an illustration of its architecture can be found in Fig. 2 in \cite{arxiv_lin_2022}.

\subsection{Input Data Strategy}
\label{sec:input_data_strategy}

Color Doppler echocardiography produces two types of information: \emph{i)} Doppler velocity, which can be corrupted by aliasing in regions of high blood speed, and \emph{ii)} Doppler power, which provides insight into the regions where velocity measurements are reliable. Using both Doppler power and velocity as input to DL models allowed them to learn how to limit the dealiasing process in regions of interest and identify ambiguous areas. Therefore, we performed an ablation study to evaluate the potential improvement provided by Doppler power. This study was conducted using the \nnunet architecture, known for its stability in training and optimal configurations. Specifically, three \nnunet models were trained with three combinations of input data: 1) \nnunet \#1 trained with Doppler velocity only, 2) \nnunet \#2 trained with the concatenation of Doppler velocity with Doppler power, and 3) \nnunet \#3 trained with the multiplication of Doppler velocity by Doppler power. After determining the best candidate for the input data, we trained the three models, \pdnet, \nnunet, and \trans, using this input combination to compute the Nyquist numbers $n_N$ \eqref{eq:Nyquist_number} from which the unwrapped Doppler velocities $V_u$ \eqref{eq:unwrapping} were derived. Our goal was to increment or decrement the Doppler velocities by $2n_NV_N$, not to modify them by smoothing, for example.

\subsection{Artificial Aliasing Augmentation Strategy}
\label{sec:data_augmentation_strategy}

\begin{figure}[tp]
\centering
    \includegraphics[width=0.45\textwidth]{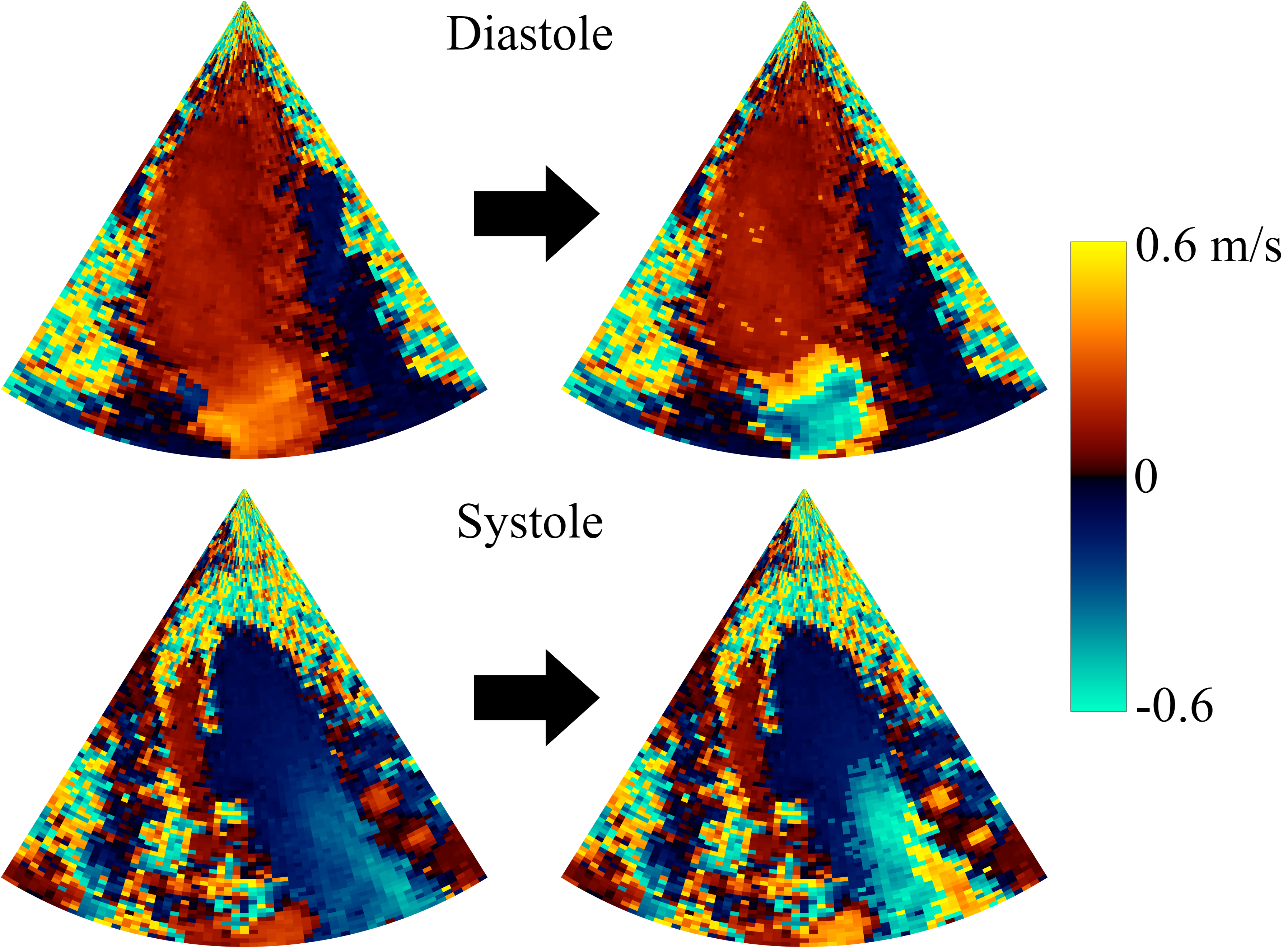}
    \caption{Generation of synthetic images with artificial aliasing artifacts (right column) from non-aliased images (left column).}
    \label{fig:realistic_aliasing_examples}
\end{figure}

Color Doppler images may exhibit aliasing only in localized regions or frames, resulting in datasets that are often imbalanced, with most pixels belonging to the background class (\ie, without aliasing). To address this issue, we used standard data augmentation techniques such as rotation, flipping etc., during training. We also proposed an additional data augmentation technique, which we called artificial aliasing augmentation, to improve the generalizability of our algorithms. This technique involved identifying regions with high Doppler velocity and power on alias-free Doppler images, and applying a wrapping function defined by \eqref{eq:wrapping} with a lower Nyquist velocity to create artificial aliasing artifacts, followed by a normalization. The ground-truth references of these artificially aliased frames were created on the fly by comparing the Doppler velocities before and after this augmentation. By creating realistic artificially aliased frames, as shown in Fig. \ref{fig:realistic_aliasing_examples}, this strategy enabled us to balance the classes in training batches. To evaluate the potential benefits of artificial aliasing augmentation, we conducted an additional ablation study where we tested the three DL models with and without this technique during training.

\subsection{\DeAN: State-of-the-Art non DL-Based Dealiasing Method for CDI}
\label{sec:sota_methods}

\DeAN, mentioned in Section \ref{sec:introduction}, is currently one of the most powerful non DL-based methods for color Doppler dealiasing. It is a semi-supervised method with a hyperparameter, $Q$. To unwrap aliased pixels, \DeAN first segments color Doppler images using a region-merging scheme based on the Hoeffding's probability inequality. Then, \DeAN compares each segmented region with its nearest neighbors and performs dealiasing if necessary. This step is based on the assumption that the largest segment is not aliased and is repeated until all the segments have been analyzed. The main drawback of this method is the need to manually search for the optimal $Q$ hyperparameter for each frame to obtain the best dealiasing results. We compared the dealiasing performance of \DeAN with both the default $Q=10$ and with the manually optimized $Q$ hyperparameter, against the three DL methods.

\subsection{Evaluation metrics}
\label{sec:evaluation_metrics}

All three DL models were designed to output the Nyquist numbers $n_N$; the dealiased velocity maps, $V_u$, were recovered using \eqref{eq:unwrapping}. To evaluate the accuracy of the dealiased velocity maps and the Nyquist numbers outputted by each method, we computed four evaluation metrics.

We compared the dealiased Doppler velocity maps $V_u$ with the ground-truth alias-free Doppler velocity maps $V_{ref}$ by computing the cosine similarity index:
\begin{equation}
    \text{Cosim}(V_u,V_{ref})= \frac{V_u \cdot V_{ref}}{\lVert  V_u \rVert \cdot \lVert  V_{ref} \rVert}
\end{equation}
Cosine similarity is a commonly used similarity measure for comparing text data or images. We used this similarity index in our previous work on color Doppler dealiasing \cite{media_muth_2011}. In addition, we computed three classification metrics to verify whether each pixel was classified correctly on color Doppler images. The first classification metric was the balanced accuracy score, which is more suitable for unbalanced datasets than the classical accuracy score. It was calculated using the following formula:
\begin{equation}
    \text{Accuracy} = \frac{1}{2} \times \left( \frac{\text{TP}}{\text{TP}+\text{FN}} + \frac{\text{TN}}{\text{TN}+\text{FP}} \right)
\end{equation}
where TP, FN, TN and FP refer to true positives, false negatives, true negatives, and false positives, respectively. Besides, the classical recall ($\frac{\text{TP}}{\text{TP}+\text{FN}}$) and precision ($\frac{\text{TP}}{\text{TP}+\text{FP}}$) metrics were also computed to evaluate the overall performance of the methods. 

To ensure the reliability and relevance of the results, we conducted a 9-fold cross-validation to compute the scores presented in each table in Section \ref{sec:experimental_results}. For each fold, we split the dataset into training, validation, and test sets using a ratio of 36/4/5 patients. This resulted in an average of 2,974, 330, and 413 color Doppler frames for the training, validation, and test sets, respectively.

\section{Experiment Setup and Results}
\label{sec:experimental_setup_results}

\subsection{Dataset and Training Strategies}
\label{sec:dataset}

\begin{figure*}[tp]
    \centering
    \includegraphics[width=0.8\textwidth]{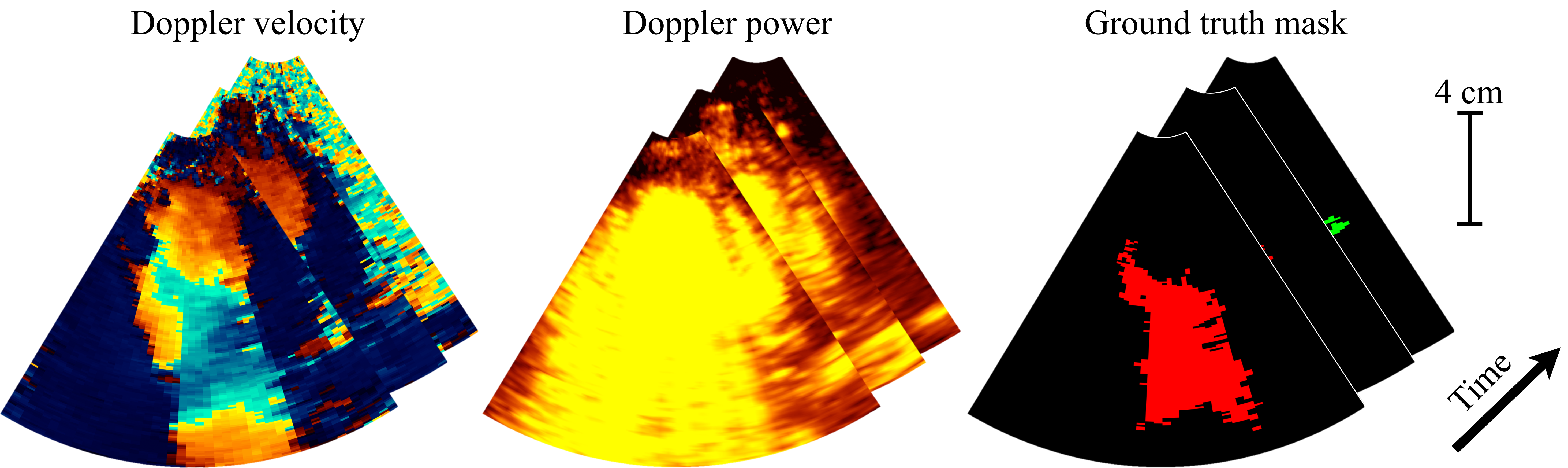}
    \caption{Generation of the ground-truth masks from the Doppler velocities. The red, black, and green segmentation masks correspond to Nyquist numbers $n_N=1$, $0$, and $-1$, respectively.}
    \label{fig:dataset}
\end{figure*}

\subsubsection{Color Doppler dataset}
\label{sec:color_doppler_dataset}
To evaluate the performance of our methods, we used a color Doppler echocardiographic dataset of 45 patients that were acquired using a Vivid 7 ultrasound system (GE Healthcare, USA) and a GE 5S cardiac sector probe (bandwidth = 2-5 MHz). Doppler velocity and power data prior to scan conversion were extracted into HDF formats using EchoPAC software (GE Healthcare). The EchoPAC software returned unitless power data in the range of 1-100. The power data, $P$, were compressed by taking the logarithm and then scaled to [0,1]: $log(P)/2 \leftarrow P$. The cardiologist used the default settings (center frequency, pulse length, pulse repetition frequency, packet size, clutter filter, etc.). These proprietary parameters are masked and could not be extracted. In most cases, the Nyquist velocity was about 0.6 m/s. Assuming a center frequency of 3 MHz, the pulse repetition frequency was approximately 4500 Hz. These anonymized data came from a previous published study \cite{mehregan_doppler_2014}. The patients were examined in the echocardiography laboratory under standard medical conditions. As a result, some patients had significant heart disease, while others had no visible pathology. The random selection of patients and their anonymization prevent us from knowing their demographic and pathological status. The sequences were acquired in the apical three-chamber view and included both Doppler velocity and power information. Each sequence covered at least one complete cardiac cycle, resulting in a total of 1,338 aliased frames and 2,379 non-aliased frames. Since color Doppler has a relatively low frame rate of 10 to 15 frames per second in clinical echocardiography, we considered each frame to be independent. To avoid interpolation artifacts, the data were collected and processed in polar coordinates (i.e., before scan-conversion), but for better visualization, the results were presented in Cartesian coordinates.

The training, validation, and test data sets, i.e., the pairs of original and alias-free Doppler velocity maps, were generated by an experienced analyzer. For this task, the non-scan-converted clinical Doppler maps were oversegmented and labeled using a statistical growing region method (see Fig. 1.B in \cite{media_muth_2011}). The analyst manually identified the aliased regions, specifically focusing on those related to intraventricular blood flow, which were then corrected by applying $\pm 2 V_N$. The noisy regions associated with low Doppler power were left unchanged. Examples of color Doppler frames along with their reference segmentations are illustrated in Fig. \ref{fig:dataset}.

\vspace*{0.3cm}

\subsubsection{Training strategies}
\label{sec:training_strategies}

To train the deep learning (DL) methods described in Section \ref{sec:methodology}, we performed supervised learning using the ground-truth segmentations from our in-house dataset. Besides applying the data augmentation strategies mentioned in Section \ref{sec:data_augmentation_strategy}, we further addressed the class imbalance of our dataset by ensuring that each batch contained at least one aliased image, whether real or synthetic. The \trans model was designed using the official implementation proposed in its GitHub repository\footnote{https://github.com/xianlin7/BATFormer}. This model took color Doppler images resized to 256 $\times$ 256 pixels as input and was trained for 400 epochs. On the other hand, \nnunet and \pdnet were implemented using the ASCENT framework\footnote{https://github.com/creatis-myriad/ASCENT}. For these two approaches, we used a patch-wise approach to preserve the resolution of the input data. The models were trained for 1000 epochs to prevent any potential under/overfitting. More details on the \trans, \nnunet, and \pdnet architectures are provided in Table \ref{tab:network_architectures}.

\vspace*{0.3cm}

\subsection{Experimental Results}
\label{sec:experimental_results}
\subsubsection{Doppler power information was useful in dealiasing difficult case}
\label{sec:doppler_power}

Table \ref{tab:unet_doppler_power} reports the results of the ablation study aimed at identifying the optimal combination of input data. The results indicate that the three \nnunet models performed similarly across all metrics, implying that incorporating Doppler power information in the input data did not substantially improve the models' performance. However, upon evaluation on a challenging test set (right part of Table \ref{tab:unet_doppler_power}), the model that was trained with the multiplication of Doppler velocity and Doppler power (\nnunet \#3) demonstrated better performance for all metrics except precision. The last two columns of Fig. \ref{fig:dealiased_images} show two samples taken from the difficult fold, where the aliased and non-aliased regions had similar hues. This made the correction of the aliased velocities difficult. Thus, although not critical, using the Doppler velocity-Doppler power product as input data is recommended as it can enhance the models' generalization ability, especially for challenging data. For subsequent experiments and results, we trained all DL methods with this input combination.

\begin{table*}[tp]
    \caption{Dealiasing by \nnunet trained with different combinations of input data using 9-fold cross-validation. \textit{\#1}, \textit{\#2}, and \textit{\#3} correspond to the use of Doppler velocity only, the concatenation of Doppler velocity with Doppler power, and the multiplication of the Doppler velocity by Doppler power as input data, respectively. The \textit{Difficult fold} column shows the evaluation results of different \nnunet models on a challenging test set containing color Doppler frames with aliased and non-aliased regions of similar hue.}
    \centering
    \begin{tabular}{c c c c c | c c c c}
    \toprule
    \multirow{4}*{Methods} & \multicolumn{4}{c}{Full dataset (Number of frames = 3,717)} & \multicolumn{4}{c}{Difficult fold (Number of frames = 413)} \\ 
    \cmidrule(lr){2-5} \cmidrule(lr){6-9}
     & Cosim & Accuracy & Recall & Precision & Cosim & Accuracy & Recall & Precision \\ 
     & $\pm \sigma$ & $\pm \sigma$ & $\pm \sigma$ & $\pm \sigma$ & $\pm \sigma$ & $\pm \sigma$ & $\pm \sigma$ & $\pm \sigma$ \\
    \midrule
    \multirow{2}*{\begin{tabular}{c} \nnunet \#1 \\ \textit{velocity only} \end{tabular}}
    & \textbf{0.99} & \textbf{0.95} & 0.90 & \textbf{0.92} & 0.98 & 0.93 & 0.86 & \textbf{0.95} \\
     & $\pm$0.01 & $\pm$0.03 & $\pm$0.07 & $\pm$0.06 & $\pm$0.01 & $\pm$0.06 & $\pm$0.13 & $\pm$0.01 \\[0.3cm]
     
    \multirow{2}*{\begin{tabular}{c} \nnunet \#2 \\ \textit{velocity-power concatenation} \end{tabular}}
    & \textbf{0.99} & \textbf{0.95} & \textbf{0.91} & 0.91 & 0.98 & \textbf{0.94} & 0.88 & 0.92 \\
     & $\pm$0.01 & $\pm$0.03 & $\pm$0.07 & $\pm$0.05 & $\pm$0.01 & $\pm$0.07 & $\pm$0.13 & $\pm$0.05 \\[0.3cm]
     
    \multirow{2}*{\begin{tabular}{c} \nnunet \#3 \\ \textit{velocity-power multiplication} \end{tabular}}
    & \textbf{0.99} & \textbf{0.95} & 0.89 & 0.90 & \textbf{0.99} & \textbf{0.94} & \textbf{0.89} & 0.92 \\
     & $\pm$0.01 & $\pm$0.06 & $\pm$0.13 & $\pm$0.11 & $\pm$0.01 & $\pm$0.07 & $\pm$0.14 & $\pm$0.02 \\      
    \bottomrule
    \end{tabular}
    \label{tab:unet_doppler_power}
\end{table*}

\vspace*{0.3cm}

\subsubsection{\pdnet outperformed its original counterpart}
\label{sec:pdnet_results}

We conducted a study to determine the optimal number of iterations for updating the primal and dual variables in both \oripdnet and \pdnet, given the sensitivity of this type of method to this parameter. The results are shown in Table \ref{tab:pdnet_tab}. From this table, we can see that \oripdnet reached a plateau after 10 iterations, and beyond 20 iterations, it became unstable during training and failed to produce results. On the other hand, \pdnet exhibited greater training stability and reached a performance plateau after 20 iterations. Moreover, \pdnet achieved better optimal results than \oripdnet for all metrics except for the cosine similarity index, where both methods performed equally well. These results suggest that using the same feature maps in the primal-dual approach is more suitable for the dealiasing task and support the use of our deep unfolding network. Additionally, it is worth mentioning that \pdnet had only 30,000 parameters, making it the lightest of the three DL models tested, as detailed in Table \ref{tab:network_architectures}. The significant reduction in parameters of \pdnet as compared with other models was due to the inclusion of the forward operator as prior information and the use of the same feature maps per iteration.

\begin{table*}[tp]
    \caption{9-fold cross-validation dealiasing results of \pdnet trained with different number of iterations (\textit{\# iter.}) for updating the primal and dual variables. The results on the left were obtained with \oripdnet, \ie, different feature maps per iteration, while the results on the right were obtained with the proposed \pdnet using the same feature maps for each iteration.}
    \centering
    \begin{tabular}{c c c c c | c c c c}
    \toprule
    \multirow{4}*{\# iter.} & \multicolumn{4}{c}{\oripdnet: Different feature maps per iteration} & \multicolumn{4}{c}{\pdnet: Same feature maps for each iteration} \\ 
    \cmidrule(lr){2-5} \cmidrule(lr){6-9}
     & Cosim & Accuracy & Recall & Precision & Cosim & Accuracy & Recall & Precision \\ 
     & $\pm \sigma$ & $\pm \sigma$ & $\pm \sigma$ & $\pm \sigma$ & $\pm \sigma$ & $\pm \sigma$ & $\pm \sigma$ & $\pm \sigma$ \\
    \midrule
    \multirow{2}*{1} & 0.95 & 0.86 & 0.71 & 0.51 & 0.95 & 0.87 & 0.74 & 0.52 \\
     & $\pm$0.03 & $\pm$0.06 & $\pm$0.13 & $\pm$0.09 & $\pm$0.03 & $\pm$0.06 & $\pm$0.12 & $\pm$0.10 \\[0.3cm] 
    \multirow{2}*{10} & \textbf{0.98} & \textbf{0.92} & \textbf{0.84} & 0.78 & \textbf{0.98} & 0.92 & 0.84 & 0.80 \\
     & $\pm$0.02 & $\pm$0.06 & $\pm$0.13 & $\pm$0.11 & $\pm$0.02 & $\pm$0.06 & $\pm$0.12 & $\pm$0.10 \\[0.3cm]
     \multirow{2}*{20} & \textbf{0.98} & \textbf{0.92} & \textbf{0.84} & \textbf{0.80} & \textbf{0.98} & \textbf{0.94} & \textbf{0.87} & \textbf{0.83} \\
     & $\pm$0.02 & $\pm$0.06 & $\pm$0.12 & $\pm$0.11 & $\pm$0.01 & $\pm$0.06 & $\pm$0.13 & $\pm$0.11 \\[0.3cm]
     \multirow{2}*{30} & \multirow{2}*{-} & \multirow{2}*{-} & \multirow{2}*{-} & \multirow{2}*{-} & 0.97 & 0.91 & 0.82 & 0.77 \\
     &  &  &  &  & $\pm$0.02 & $\pm$0.07 & $\pm$0.13 & $\pm$0.14 \\
    \bottomrule
    \end{tabular}
    \label{tab:pdnet_tab}
\end{table*}

\vspace*{0.3cm}

\subsubsection{Artificial aliasing augmentation improved the performance of segmentation-based networks}
\label{sec:artificial_aliasing_results}

The results presented in Table \ref{tab:artificial_artifacts} show that the use of artificial aliasing augmentation during training had varying effects on the performance of the different DL models. For \nnunet, there was a slight improvement in all metrics except precision. In contrast, \trans showed significant improvement in accuracy, recall, and precision metrics, with values increasing from 0.88, 0.76, and 0.85 to 0.91, 0.81, and 0.91, respectively. However, for \pdnet, the use of artificial aliasing augmentation resulted in degraded performance, with accuracy and recall decreasing from 0.94 and 0.87 to 0.88 and 0.77, respectively. These results highlight the challenge that the primal-dual-based regression methods face when generalizing to different types of aliasing. Based on the findings shown in Tables \ref{tab:unet_doppler_power} to \ref{tab:artificial_artifacts}, we determined that the multiplication of Doppler velocity and power should be used as input for all methods, while artificial aliasing augmentation should be applied during training only for the segmentation-based techniques, \ie, \nnunet and \trans.
    
\begin{table*}[tp]
    \caption{9-fold cross-validation dealiasing results of the three implemented deep learning solutions trained with and without the proposed artificial aliasing augmentation strategy.}
    \centering
    \begin{tabular}{c c c c c | c c c c}
    \toprule
    \multirow{4}*{Methods} & \multicolumn{4}{c}{Without artificial aliasing augmentation} & \multicolumn{4}{c}{With artificial aliasing augmentation} \\ 
    \cmidrule(lr){2-5} \cmidrule(lr){6-9}
     & Cosim & Accuracy & Recall & Precision & Cosim & Accuracy & Recall & Precision \\ 
     & $\pm \sigma$ & $\pm \sigma$ & $\pm \sigma$ & $\pm \sigma$ & $\pm \sigma$ & $\pm \sigma$ & $\pm \sigma$ & $\pm \sigma$ \\
    \midrule
    \multirow{2}*{\pdnet} & \textbf{0.98} & \textbf{0.94} & \textbf{0.87} & 0.83 & \textbf{0.98} & 0.88 & 0.77 & \textbf{0.84} \\
     & $\pm$0.02 & $\pm$0.06 & $\pm$0.12 & $\pm$0.10 & $\pm$0.01 & $\pm$0.08 & $\pm$0.16 & $\pm$0.09 \\[0.3cm]
    \multirow{2}*{\nnunet} &\textbf{0.99} & 0.95 & 0.89 & \textbf{0.90} & \textbf{0.99} & \textbf{0.96} & \textbf{0.91} & 0.89 \\
     & $\pm$0.01 & $\pm$0.03 & $\pm$0.09 & $\pm$0.06 & $\pm$0.01 & $\pm$0.03 & $\pm$0.06 & $\pm$0.06 \\[0.3cm] 
    \multirow{2}*{\trans} & \textbf{0.98} & 0.88 & 0.76 & 0.85 & \textbf{0.98} & \textbf{0.91} & \textbf{0.81} & \textbf{0.91} \\
     & $\pm$0.02 & $\pm$0.08 & $\pm$0.17 & $\pm$0.11 & $\pm$0.02 & $\pm$0.07 & $\pm$0.14 & $\pm$0.07 \\    
    \bottomrule
    \end{tabular}
    \label{tab:artificial_artifacts}
\end{table*}  

\vspace*{0.3cm}

\subsubsection{\nnunet gave the best dealiasing results}
\label{sec:nnunet_results}

Table \ref{tab:final_results} presents the final results of our study, where we compare the three DL methods with their optimal configurations against the \DeAN algorithm. We observe that all three DL methods outperformed the \DeAN algorithm, even the version with the manually chosen optimal $Q$ hyperparameter. This outcome confirms the potential of DL methods for color Doppler dealiasing. Among the DL methods, \nnunet achieved the highest scores overall, with a cosine similarity close to 1, an accuracy of 0.96, a recall of 0.91, and a precision of 0.89. Therefore, we conclude that \nnunet is the best DL approach currently available for dealiasing tasks in echocardiography. Additionally, it is interesting to note that \trans showed a clear improvement when we increased the amount of synthetic data, indicating that this type of approach requires a larger dataset to improve its performance for the dealiasing task. Finally, it is worth noting that \pdnet achieved promising results with 233 times fewer parameters compared to \nnunet, highlighting the potential of incorporating analytical context into the DL framework to regularize the solution space.

\begin{figure*}[tp]
    \centering
    \includegraphics[width=0.75\textwidth]{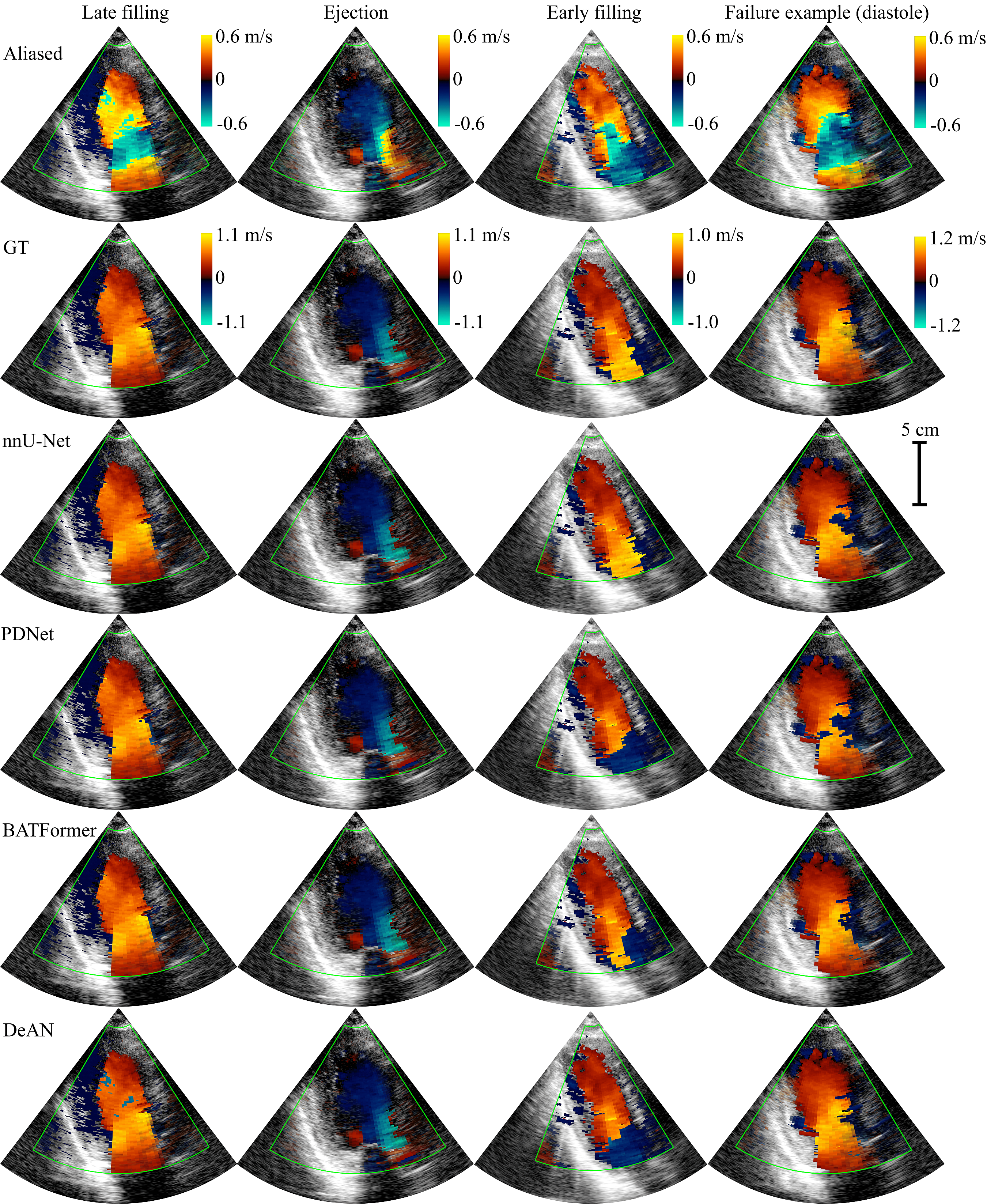}
    \caption{Color Doppler images acquired during: from left to right column, late diastolic filling, systolic ejection, early diastolic filling, and diastole (a failed case). They were dealiased by DL-based methods and by \DeAN with optimized $Q$ hyperparameters. First row: aliased raw color Doppler. Second row: alias-free ground truth (GT).}
    \label{fig:dealiased_images}
\end{figure*}

We also provide a visual inspection of the performance of the various methods on aliased images with different degrees of difficulty in Fig. \ref{fig:dealiased_images}. We can see that the DL methods performed similarly well on the easy and moderate cases (first two columns), but \nnunet produced the closest results to the reference on the more challenging case (third column). This finding is consistent with the quantitative results presented in Table \ref{tab:final_results}. The last column in Fig. \ref{fig:dealiased_images} shows an example where no method was able to handle aliasing correctly. This example is similar to the one in the first column, but with a more pronounced level of aliasing. In this particular case, the \DeAN method gave the best results. This suggests that it would be advisable to supplement the training set with challenging configurations.

\begin{table}[tp]
    \caption{9-fold cross-validation final dealiasing results of DL methods with their best configurations against non-DL \DeAN method.}
    \centering
    \begin{tabular}{c c c c c}
    \toprule
    \multirow{2}*{Methods} & Cosim & Accuracy & Recall & Precision \\ 
     & $\pm \sigma$ & $\pm \sigma$ & $\pm \sigma$ & $\pm \sigma$ \\
    \midrule
\multirow{2}*{DeAN (Q=10)} & 0.95 & 0.81 & 0.62 & 0.53 \\
     & $\pm$0.03 & $\pm$0.08 & $\pm$0.16 & $\pm$0.20 \\[0.3cm]
    \multirow{2}*{DeAN (Optimized Q)} & 0.98 & 0.91 & 0.83 & 0.80 \\
     & $\pm$0.01 & $\pm$0.04 & $\pm$0.08 & $\pm$0.13 \\ [0.1cm]
    \midrule
    \multirow{2}*{\pdnet} & 0.98 & 0.94 & 0.87 & 0.83 \\
     & $\pm$0.02 & $\pm$0.06 & $\pm$0.12 & $\pm$0.10 \\ [0.3cm]
    \multirow{2}*{\nnunet} & \textbf{0.99} & \textbf{0.96} & \textbf{0.91} & 0.89 \\
     & $\pm$0.01 & $\pm$0.03 & $\pm$0.06 & $\pm$0.06 \\[0.3cm] 
    \multirow{2}*{\trans} & 0.98 & 0.91 & 0.81 & \textbf{0.91} \\
     & $\pm$0.02 & $\pm$0.07 & $\pm$0.14 & $\pm$0.07 \\ 
    \bottomrule
    \end{tabular}
    \label{tab:final_results}
\end{table}  

\section{Discussions}
\label{sec:discussions}

Color Doppler imaging takes high-pass-filtered in-phase/quadrature (I/Q) data of the same region of interest acquired along the slow-time axis and differentiates them pairwise using a lag-1 autocorrelator. The resulting maps show blood displacements between two consecutive slow-time samples. By its very nature, color Doppler imaging is an interferometric technique that enables the measurement of displacements with a precision that can reach fractions of the center wavelength. Similarly, synthetic aperture radar interferometry (InSAR), a remote sensing technique used to map the Earth's surface deformations, generates interferograms that display ground-surface displacements. Like color Doppler, most interferometric imaging techniques in fields such as medical imaging, remote sensing, and optical metrology (e.g., phase-contrast MRI, InSAR, holographic interferometry) are subject to aliasing, i.e., jumps that occur whenever the phase shift equals $\pm\pi$. Our study aimed to address the issue of phase jumps. Among the traditional methods for phase unwrapping, one can mention: \textit{i)} graph cuts \cite{dong2017,bioucas2007}, which involve representing the wrapped phase data as a graph and determining the minimum cut that separates the known and unknown phase values; \textit{ii)} least-squares approaches, which minimize the differences between partial derivatives of the wrapped phase and those of the unwrapped solution \cite{hunt1979,josaa_ghiglia_1994}. Specifically for color Doppler echocardiography, Muth \etal developed \DeAN, a dealiasing algorithm based on statistical region merging \cite{media_muth_2011}, which was used in this study for comparative purposes.

Recently, deep learning (DL) techniques have been used to improve traditional methods in phase unwrapping \cite{spoorthi2018, wang2019, uffc_nahas_2020, zhou2021}. Our goal was to obtain alias-free color Doppler echocardiography by applying DL to dealias clinical Doppler velocity fields. DL-based approaches have been introduced for 2D phase unwrapping in InSAR \cite{zhou2021}. Unlike echocardiographic images, InSAR interferograms are subject to multiple wraps, making the networks proposed in our study not suited since we focused only on single aliasing. On the other hand, InSAR images are not subject to significant clutter, whereas clutter in Doppler echocardiography generates substantial noise near moving tissues, making 2D phase unwrapping challenging. As a result, non-DL approaches such as graph cuts or least-squares methods, which work well for InSAR interferograms, are not effective for echocardiographic Doppler fields. Although the \DeAN technique largely solved the problem, it still fails in some situations, as shown in our study. Therefore, we turned to DL and conducted an in-depth analysis and comparison of three architectures, including \pdnet, which utilizes an unfolding framework. In addition, we illustrated the potential benefits of incorporating Doppler power information, since low power generally indicates poor blood Doppler signal. To better balance the aliased and non-aliased input data during training, we resorted to data augmentation by generating synthetic aliasing.

\subsection{Comparison of the DL Methods}
\label{sec:discussion_DL}
Our study found that the three DL methods we tested (\pdnet, \nnunet, and \trans) outperformed the non-DL \DeAN method for color Doppler dealiasing. Notably, we observed that \nnunet had the best performance, suggesting that the 2D U-Net architecture used in \nnunet may be particularly well-suited for this task due to its ability to effectively capture spatial features.
For example, in a challenging case where the aliased and non-aliased regions had similar hues (Fig. \ref{fig:dealiased_images}, third column), \nnunet was able to unwrap correctly while other DL methods failed or were less successful. Similar structures corrected by an expert were part of the training dataset, which implies that \nnunet probably learned the flow
patterns and leveraged this knowledge to achieve successful outcomes. \pdnet also performed well, requiring $>$ 200 times fewer parameters than \nnunet, highlighting the potential for simpler DL models to achieve competitive results in color Doppler processing. Further exploration of this type of unfolding approach, including more complex modeling of the forward operator, is needed.

Although the third input strategy (velocity-power multiplication) contained less information than the second (velocity-power concatenation), it performed slightly better in the difficult fold (last row of Table \ref{tab:unet_doppler_power}). The multiplication strategy largely suppressed velocity discontinuities in noisy regions, making the training task easier. In contrast, the concatenation of Doppler power and velocity allowed the model to learn the best strategy for combining these two inputs, which could be beneficial in larger datasets.

While we did not observe significant performance gains with \trans, the addition of synthetic data improved the outcomes, indicating that \trans also has potential for color Doppler dealiasing, especially when more data is available. Our results demonstrate that DL methods can significantly improve upon traditional methods for color Doppler processing. Additionally, they underscore the importance of further investigating the performance of different DL architectures for this task and finding ways to effectively exploit the strengths of each architecture.

\subsection{Limitations and Perspectives}
\label{sec:discussion_limitations}

Color Doppler aliasing in the left ventricle mainly occurs in the mitral jet during early and late filling, as well as during ejection into the ventricular outflow tract. As depicted in the figures, the aliasing in our study was single. However, in certain valvular diseases, such as mitral stenosis or aortic regurgitation, multiple aliasing can occur in the intraventricular cavity due to the high fluctuating velocities of the turbulent jet. The nature of multiple aliasing in this context differs from that observed in InSAR, requiring specific studies to assess the feasibility of removing aliasing in areas with significant local flow perturbations. Although this remains to be verified, it is likely that a similar strategy could also work with disturbed flows, provided that we have access to Doppler data with their alias-free references. Such ground truths could be obtained by supervised correction, as in this study, and by simulations \cite{sun2022}.

Since we used a clinical ultrasound system with a color Doppler rate of 10 to 15 frames per second, our study did not exploit temporal information. In the context of high-frame-rate echocardiography \cite{faurie2016}, neural networks with enforced temporal consistency \cite{painchaud2022} or 3D \mbox{U-Net} could potentially improve dealiasing performance by leveraging temporal information. This approach would be especially relevant as high-frame-rate color Doppler is subject to more noise related to clutter signals.

\subsection{Applications in Quantitative Color Doppler}
\label{sec:discussion_quantitative}

Once corrected, color Doppler images can be used to visualize and quantify intracardiac blood flow. As mentioned in the introduction, intraventricular vector flow mapping (\textit{i}VFM) is an approach to obtain comprehensive flow information, from which hemodynamic parameters can be estimated. Using a color M-mode, it is also possible to estimate the pressure difference between the apex and the mitral base, which reflects the cardiac filling \cite{sorensen2023, hodzic2020}. However, prior dealiasing is required for this method \cite{hodzic2020}. To this end, the approaches outlined in this study could be used with color M-mode images. In a more ambitious perspective, it would be conceivable to develop neural networks that can directly infer velocity vector fields or relative pressure fields from color Doppler images, once properly trained. In this case, the dealiasing process would be intrinsically integrated into the network. The main difficulty would lie in obtaining paired input data that provide the reference values. Simulations combining flows and acoustics could provide a relevant avenue for this purpose \cite{shahriari2018, sun2022}.

\section{Conclusion}
\label{sec:conclusion}

While traditional methods are effective for interferometric imaging, they are limited for color Doppler echocardiography due to high noise generated by clutter signals in moving tissues surrounding the blood flow. We have demonstrated that deep learning (DL) techniques can achieve alias-free color Doppler echocardiography. Our proposed DL methods outperformed the non-DL \DeAN method, with \nnunet achieving the best performance, followed by \pdnet. In addition, the incorporation of power information and artificial aliasing augmentation improved the results. The application of DL techniques to color Doppler echocardiography is a promising approach that could enhance the clinical utility of this widely used imaging modality.

\bibliographystyle{IEEEtran}
\bibliography{tuffc_ling_2023}
\end{document}